\documentclass[sigconf]{acmart}

\AtBeginDocument{%
  \providecommand\BibTeX{{%
    \normalfont B\kern-0.5em{\scshape i\kern-0.25em b}\kern-0.8em\TeX}}}

\setcopyright{acmcopyright}
\copyrightyear{2024}
\acmYear{2024}
\acmDOI{XXXXXXX.XXXXXXX}

\acmConference[SEAMS 2024]{19th International Conference on Software Engineering for Adaptive and Self-Managing Systems}{April 2024}{Lisbon, Portugal}

\usepackage{xspace}
\usepackage{amsmath}

\usepackage{amssymb,amsfonts}

\usepackage{algorithmic}
\usepackage{graphicx}
\usepackage{textcomp}
\usepackage{xcolor}

\usepackage{caption}
\usepackage{subcaption}
\usepackage{url}
\usepackage{enumitem}
\usepackage{booktabs}
\usepackage{array}
\usepackage{quoting}
\usepackage{hyperref}
\usepackage{multirow}
\usepackage{makecell}
\newcolumntype{L}[1]{>{\raggedright\let\newline\\\arraybackslash\hspace{0pt}}m{#1}}
\newcolumntype{C}[1]{>{\centering\let\newline\\\arraybackslash\hspace{0pt}}m{#1}}
\newcolumntype{R}[1]{>{\raggedleft\let\newline\\\arraybackslash\hspace{0pt}}m{#1}}

\newcommand{\repourl}{\url{https://github.com/carwehlm/PARLEY}}

\newcommand{\UMC}{\textsc{URC}\xspace}%uncertainty reduction controller
\newcommand{\UUC}{\textsc{UAC}\xspace}%uncertainty aware controller
\newcommand{\ACRONYM}{\textsc{Parley}\xspace}
\usepackage[many]{tcolorbox}   
\newtcolorbox{boxA}{
    colback = white,
    boxrule = 1pt,
    colframe = black, % frame color
    boxsep=1pt, 
    left=2pt,
    right=2pt,
    top=2pt,
    bottom=2pt
}

\usepackage{listings}
\definecolor{antlrmagenta}{rgb}{1, 0, 0.40}
\lstdefinestyle{ANTLR}{
    basicstyle=\small\ttfamily\color{antlrmagenta},%
    breaklines=true,%                                      allow line breaks
    moredelim=[s][\color{green!50!black}\ttfamily]{`}{'},% single quotes in green
    moredelim=*[s][\color{black}\ttfamily]{options}{\}},%  options in black (until trailing })
    commentstyle={\color{gray}\itshape},%                  gray italics for comments
    morecomment=[l]{//},%                                  define // comment
    emph={%
        STRING, color, icon, meta, name, private, var, grammar, fragment%                                            literal strings listed here
        },emphstyle={\color{blue}\ttfamily},%              and formatted in blue
    alsoletter={:,|,;},%
    morekeywords={:,|,;},%                                 define the special characters
    keywordstyle={\color{black}},%                         and format them in black
    	tabsize=3,%
    captionpos=b,%
    keepspaces,%
}
\definecolor{prismgreen}{rgb}{0, 0.5, 0} 
\lstdefinelanguage{Prism}{ % syntax highlight via font 
        basicstyle=\color{red}\scriptsize\ttfamily, % small true type font (like courier)
        breaklines=true,%                                      allow line breaks
        keywords={bool,C,ceil,const,ctmc,double,dtmc,endinit,endmodule,endrewards, endsystem,F,false,floor,formula,G,global,I,init,int,label,max,mdp,min, module,nondeterministic,P,Pmin,Pmax,prob,probabilistic,R,rate,rewards, Rmin,Rmax,S,stochastic,system,true,U,X},% 
        keywordstyle={\bfseries\color{black}},%
        numberstyle=\tiny\color{black},%
        comment=[l] {//}, morecomment=[s]{/*}{*/}, % single and multi-line 
        commentstyle= \color{prismgreen}, % dark green 
        tabsize=2, % tab treatment (going to be fixed in Prism) 
        captionpos=b, % put captions at the bottom 
        escapechar=@ % write LaTeX comments escaped by @ symbol 
} 

\begin{document}

\title{Formal Synthesis of Uncertainty Reduction Controllers}

\author{Marc Carwehl}
\email{carwehl@cs.hu-berlin.de}
\orcid{0000-0003-0631-365X}
\affiliation{%
  \institution{\textit{Institut f\"ur Informatik, Humboldt-Universit\"at zu Berlin}}
  \streetaddress{Unter den Linden 6}
  \city{Berlin}
  \country{Germany}
  \postcode{10099}
}

\author{Calum Imrie}
\email{calum.imrie@york.ac.uk}
\orcid{0009-0004-3198-9226}
\affiliation{%
  \institution{\textit{Department of Computer Science, University of York}}
  \streetaddress{}
  \city{York}
  \country{UK}
  \postcode{}
}

\author{Thomas Vogel}
\email{thomas.vogel@cs.hu-berlin.de}
\orcid{0000-0002-7127-352X}
\affiliation{%
  \institution{\textit{Institut f\"ur Informatik, Humboldt-Universit\"at zu Berlin}}
  \streetaddress{Unter den Linden 6}
  \city{Berlin}
  \country{Germany}
  \postcode{10099}
}

\author{Genaína Rodrigues}
\email{genaina@unb.br}
\orcid{0000-0003-1661-8131}
\affiliation{%
  \institution{\textit{Department of Computer Science, University of Brasilia}}
  \streetaddress{}
  \city{Brasilia}
  \country{Brazil}
  \postcode{}
}

\author{Radu Calinescu}
\email{radu.calinescu@york.ac.uk}
\orcid{0000-0002-2678-9260}
\affiliation{%
  \institution{\textit{Department of Computer Science, University of York}}
  \streetaddress{}
  \city{York}
  \country{UK}
  \postcode{}
}

\author{Lars Grunske}
\email{grunske@cs.hu-berlin.de}
\orcid{0000-0002-8747-3745}
\affiliation{%
  \institution{\textit{Institut f\"ur Informatik, Humboldt-Universit\"at zu Berlin}}
  \streetaddress{Unter den Linden 6}
  \city{Berlin}
  \country{Germany}
  \postcode{10099}
}

\renewcommand{\shortauthors}{Carwehl, et al.}

\begin{abstract}
In its quest for approaches to taming uncertainty in self-adaptive systems (SAS), the research community has largely focused on solutions that adapt the SAS architecture or behaviour in response to uncertainty. By comparison, solutions that reduce the uncertainty affecting SAS (other than through the blanket monitoring of their components and environment) remain underexplored. Our paper proposes a more nuanced, adaptive approach to SAS uncertainty reduction. To that end, we introduce a SAS architecture comprising an \emph{uncertainty reduction controller} that drives the adaptive acquisition of new information within the SAS adaptation loop, and a tool-supported method that uses probabilistic model checking to synthesise such controllers. The controllers generated by our method deliver optimal trade-offs between SAS uncertainty reduction benefits and new information acquisition costs. We illustrate the use and evaluate the effectiveness of our approach for mobile robot navigation and server infrastructure management SAS. 
\end{abstract}

\begin{CCSXML}
<ccs2012>
   <concept>
       <concept_id>10010520.10010575.10010577</concept_id>
       <concept_desc>Computer systems organization~Reliability</concept_desc>
       <concept_significance>500</concept_significance>
       </concept>
   <concept>
       <concept_id>10010520.10010553.10010554.10010557</concept_id>
       <concept_desc>Computer systems organization~Robotic autonomy</concept_desc>
       <concept_significance>500</concept_significance>
       </concept>
   <concept>
       <concept_id>10011007.10010940.10010971.10010972.10010974</concept_id>
       <concept_desc>Software and its engineering~Layered systems</concept_desc>
       <concept_significance>500</concept_significance>
       </concept>
   <concept>
       <concept_id>10011007.10010940.10010971.10010980.10010984</concept_id>
       <concept_desc>Software and its engineering~Model-driven software engineering</concept_desc>
       <concept_significance>300</concept_significance>
       </concept>
   <concept>
       <concept_id>10011007.10010940.10010992.10010998</concept_id>
       <concept_desc>Software and its engineering~Formal methods</concept_desc>
       <concept_significance>500</concept_significance>
       </concept>
 </ccs2012>
\end{CCSXML}

\ccsdesc[500]{Computer systems organization~Reliability}
\ccsdesc[500]{Computer systems organization~Robotic autonomy}
\ccsdesc[500]{Software and its engineering~Layered systems}
\ccsdesc[300]{Software and its engineering~Model-driven software engineering}
\ccsdesc[500]{Software and its engineering~Formal methods}

\keywords{controller synthesis, uncertainty, self-adaptive systems}


\maketitle

\section{Introduction}

Essential services from all sectors of the economy and society rely on the effective operation of complex software-intensive systems. These systems range from sophisticated road traffic management software and public clouds running business-critical applications to cyber-physical systems 
from manufacturing. More often than not, they are used in real-world applications characterised by high levels of uncertainty related, for instance, to environmental changes, component failures, measuring inaccuracies, and user actions. To deliver their required functionality in such circumstances, software-intensive systems need to ``tame'' this uncertainty through \emph{self-adaptation}~\cite{EsfahaniKM11,Mahdavi-Hezavehi20}. Self-adaptation is a process that involves the use of closed-control loops to \emph{monitor} the system and its environment for relevant changes, to \emph{analyse} the impact of these changes, to \emph{plan} system adaptations that accommodate the changes, and to \emph{execute} (i.e., to implement) these adaptations. Software-intensive systems that employ such monitor-analyse-plan-execute (or `MAPE', cf.~\cite{kephart2003vision}) control loops are termed \emph{self-adaptive systems} (SAS). 

The growing demand for SAS in many application domains~\cite{weyns2022preliminary,weyns2023self} has led to the development of numerous self-adaptation solutions over the past two decades. Nevertheless, the vast majority of these solutions focus on determining and performing SAS adaptation tactics that take uncertainty into account. The complementary approach of \emph{reducing epistemic uncertainty} (i.e., the uncertainty due to insufficient knowledge)---other than through a blanket monitoring of the system and its environment in the first step of the MAPE  loop---is largely unexplored by existing SAS solutions. 

In this paper, we argue that SAS can achieve better tradeoffs between adaptation outcomes and costs by combining established uncertainty-aware adaptation solutions with an adaptive approach to epistemic uncertainty reduction. To motivate the need for our hybrid self-adaptation paradigm, we refer to SAS exemplars proposed by the SEAMS research community~\cite{SEAMS-exemplars}. 

As a first example, consider the UNDERSEA exemplar from~\cite{gerasimou2017undersea}. This is an underwater robot tasked with measuring the oceanic salinity or temperature with a given accuracy and under energy use constraints, which requires the robot to adaptively switch on/off its sensors and to vary its speed depending on environmental conditions. Assuming that the mission needs to be performed within a designated perimeter which contains obstacles such as underwater rocks, the MAPE loop controlling the robot needs to consider the robot's position in its decision-making. However, because of variable underwater currents that are difficult to model, this position is affected by epistemic uncertainty whose resolution requires the robot to navigate to the sea surface for GPS access. This uncertainty-reduction operation consumes significant time and energy. As such, it should ideally be performed adaptively (based on need), by considering factors such as the estimated distance between the robot and the nearest obstacle/perimeter boundary, robot speed, and estimated underwater current direction and speed. 

As another example, consider the TAS exemplar from~\cite{weyns2015tele}. TAS is a telehealth service-based system that uses third-party services (i)~to analyse the vital parameters of home-based patients with long-term health conditions, and, when the patient condition changes significantly, (ii)~to order new medication, or (iii)~to alert a medical team. Its MAPE loop is responsible for ensuring that the TAS reliability and response time remain within acceptable bounds, by switching to a functionally equivalent ``backup'' service when the reliability or performance of any of its three services becomes inadequate. However, because the backup services can also experience technical difficulties, the reliability and response time that the MAPE loop assumes for each of them are affected by epistemic uncertainty. To reduce this uncertainty, and thus to avoid switching to a backup service that has become unavailable or too slow, TAS should occasionally test these services by invoking their functions. Given the unavoidable overheads of these uncertainty-reduction invocations, their timing, frequency and number should be continually adapted to reflect the current TAS workload, the recent reliability and performance trends of the services used by TAS, etc.

\medskip\noindent
\textbf{Problem definition:} The UNDERSEA and TAS scenarios we described are instances of a general problem faced by SAS whose MAPE loops make decisions based on estimates of variables affected by epistemic uncertainty. The precision of these estimates can be increased by using SAS-specific uncertainty-reduction ``services'',\footnote{These services may suffer from \emph{aleatoric uncertainty}, i.e., irreducible uncertainty due, for instance, to natural variability; e.g., the GPS services used by the UNDERSEA robot from our example can only provide the robot location with a certain accuracy.} accessing these services incurs a cost that may consist of CPU or memory overheads, bandwidth or energy use, carbon footprint, etc. As such, using these services continuously is unaffordable. Moreover, invoking them with a constant frequency is likely to yield suboptimal tradeoffs between the uncertainty-reduction cost and the effectiveness of the MAPE decision-making. Thus, the \emph{adaptive uncertainty reduction problem} tackled in our paper is to determine \emph{which uncertainty-reduction service(s) should be invoked when} in order to ensure that the SAS goals are optimally satisfied. 

\begin{figure}
    \centering
    \includegraphics[width=\columnwidth]{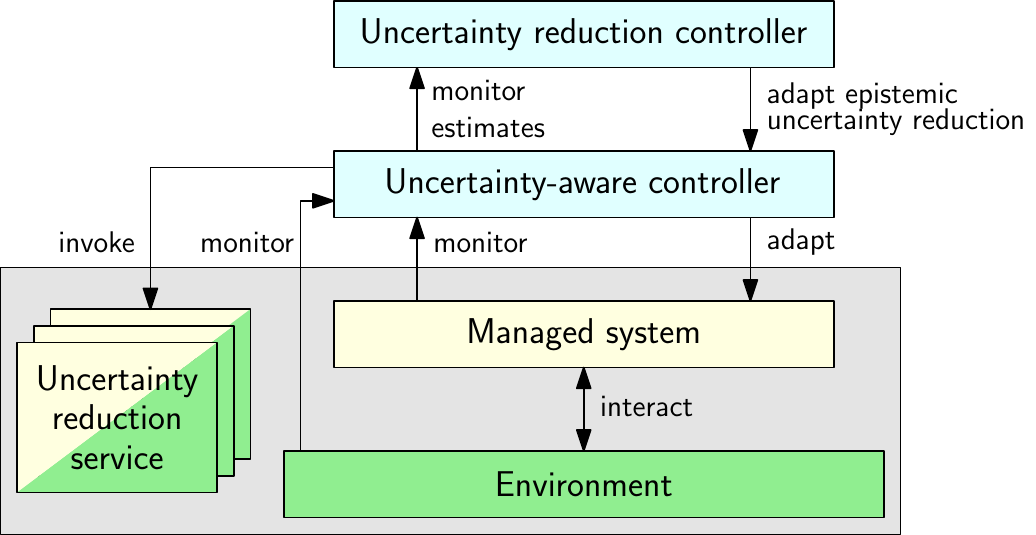}
    \vspace{-2em}
    \caption{\ACRONYM self-adaptive system architecture: an uncertainty reduction controller drives the adaptive reduction of epistemic uncertainty through the invocation of uncertainty reduction ``services'' provided by the managed system, its environment, or a combination thereof.}
    \label{fig:architecture}
    \vspace{-1em}
\end{figure}

\medskip\noindent
\textbf{Our approach:} To address the problem summarised above, we introduce a new paradigm for ada\underline{p}tive uncert\underline{a}inty \underline{r}eduction in se\underline{l}f-adaptiv\underline{e} s\underline{y}stems, or \ACRONYM\footnote{From the French \emph{parler} (to speak); parley means to discuss and come to an agreement.} for short. The fundamental premise behind \ACRONYM is that acquiring new knowledge to reduce epistemic uncertainty represents one of the paramount tasks that a SAS should be concerned with. In line with this premise, \ACRONYM is underpinned by a new SAS architecture (Fig.~\ref{fig:architecture}) that includes a dedicated \emph{uncertainty reduction controller} (\UMC). The role of this new type of controller is to monitor the estimated values of uncertainty-affected variables that the existing MAPE loop (referred to here as \textit{uncertainty-aware controller (UAC)}) operates with and to dynamically adapt this controller's use of the available uncertainty-reduction services by considering factors such as those from our earlier UNDERSEA and TAS scenarios. The use of different controllers for uncertainty reduction and managed-system adaptation to uncertainty in our \ACRONYM architecture has two major benefits. First, as with any architecture that promotes a separation of concerns between different system functions, \ACRONYM can lead to less complex and easier to maintain control loops than a monolithic architecture. Second, our two-tier control architecture makes the augmentation of an existing SAS with an uncertainty reduction controller straightforward.

In addition to this new architecture, \ACRONYM comes with a tool-supported method for the formal synthesis of uncertainty reduction controllers for SAS whose behaviour can be modelled using probabilistic state-transition models such as discrete-time Markov chains (DTMCs). This method uses a combination of probabilistic model checking and multi-objective genetic algorithms to synthesise URCs guaranteed to satisfy strict reliability, performance and other quality constraints, and to be Pareto-optimal with respect to a set of quality optimisation objectives.
 
\medskip\noindent
\textbf{Contributions:} The main contributions of our paper are:
\begin{itemize}
    \item The \ACRONYM hybrid self-adaptation paradigm, and associated two-tier controller architecture;
    \item The \ACRONYM method for synthesising correct-by-construction uncertainty reduction controllers;
    \item A toolchain which automates the application of the URC synthesis method, and which includes a new tool for augmenting the discrete-time Markov chain model of a SAS with the new states and transitions required for the URC synthesis;
    \item The evaluation of \ACRONYM within SAS case studies from the mobile robot navigation and server infrastructure management domains.
\end{itemize}

\medskip\noindent
\textbf{Paper structure:} The rest of the paper is organised as follows. In Section~\ref{sec:related}, we compare \ACRONYM to related research on self-adaptive systems. Next, Section~\ref{sec:running-example} provides a running example used to illustrate the components and stages of \ACRONYM in its description from Section~\ref{sec:proposal}. Finally, Section~\ref{sec:evaluation} presents our evaluation of \ACRONYM, and Section~\ref{sec:conclusion}  concludes the paper with a brief summary and suggestions for further work.

\section{Related Work \label{sec:related}}

Handling uncertainty has been one of the driving forces of research in the area of self-adaptive systems ~\cite{CalinescuMPW20, GieseBPRIWC11, Mahdavi-Hezavehi20, RamirezJC12, WhittleSBCB10, WeynsCMTABBBCDEGGYGHLLMMMRPQSVZ23}. In the following, we review selected approaches that are particularly related to \ACRONYM. 
Uncertainty handling may follow a control-theoretical ~\cite{michelmore2020uncertainty, Kobayashi:NFM21, Shevtsov+2019,solano2019taming}, an architecture-based ~\cite{CamaraSCP:2018, martins2019alphapomdp, Moreno:SEAMS18, paucar2018re, Weyns:TOSEM23} or a planing-based~\cite{Agha:IROS14, paucar2018re, martins2019alphapomdp, QosMOS,EsfahaniKM11,SBS:SoSYM22} adaptation approach. 

\medskip\noindent
\textbf{Control-theoretical approaches:}
Shevtsov et al.~\cite{Shevtsov+2019} apply control theory to deal with adaptation problems for systems with strict goals and control theoretical requirements (set point, thresholds, optimisation) as well as to handle and provide assurances under various sources of uncertainty. In contrast, Michelmore et al.~\cite{michelmore2020uncertainty} developed a framework for evaluating the safety of autonomous driving using end-to-end Bayesian Neural Network (BNN) controllers. With their work, uncertainty estimates for the controller’s decisions can be obtained with a priori statistical guarantees. While in their work they optimise for safety constraints, our optimal policy aims for a broader spectrum where epistemic uncertainties can be mitigated as the system objectives are satisfied. The work by Kobayashi et al.~\cite{Kobayashi:NFM21} proposes a methodology to build more robust controllers against perceptual uncertainty through an automated workflow. By providing optimal policies over the behaviour,  
our work goes one step further as it takes probabilities into account and provides means to build extensible controllers through the separation of concerns between uncertainty-aware and -reduction controllers. Moreover, the controller synthesis in \ACRONYM can inherently accommodate not only multiple parameters but also multiple thresholds through trading off objectives. Solano et al.~\cite{solano2019taming} propose an assurance process to handle uncertainty through a generative approach that uses a goal model augmented with uncertainties. As an outcome, reliability and cost algebraic formulae are used by a PID controller to provide policies and assure the properties of the managed system. Their solution, however, does not allow for architecturally decoupling the controllers' behaviours and concerns, which could render the controller's maintainability and scalability infeasible.

\medskip\noindent
\textbf{Architecture-based adaptation approaches:}
Regarding the handling of uncertainties following an architecture-based adaptation approach, Weyns and Iftikar~\cite{Weyns:TOSEM23} propose ActivForms-ta, an architec\-ture-based adaptation approach based on the MAPE-K reference model. Moreno et al.~\cite{Moreno:SEAMS18} introduced the concept of uncertainty reduction to manage uncertainty in self-adaptive systems and show how uncertainty reduction decisions can be integrated into self-adaptation decisions. Our work addresses some challenges put forward by Moreno et al. Besides following a MAPE-K adaptation approach, \ACRONYM focuses on allowing an explicit representation and the correct-by-construction synthesis of  
uncertainty controllers. In particular, employing the estimate of the uncertainty-aware controller, \ACRONYM can act on handling uncertainties if and when necessary. We believe the work by Camara et al. in \cite{CamaraSCP:2018} is the closest to ours. In that work, they present a formal reasoning technique based on stochastic multiplayer games to improve decision-making through uncertainty-aware and uncertainty-ignorant decision-making in regions of the state space in which aleatoric uncertainty matters. \ACRONYM also benefits from such a separation of uncertainties for the controllers together with a formal technique underpinning our uncertainty reduction service in the formalism of pDTMC. Additionally, \ACRONYM resorts to a meta-heuristic approach to find near-optimal adaptation policies. 
Kreutz et al.~\cite{kreutz2022towards} recently proposed a new approach to model uncertain knowledge to estimate the best adaptation tactic. We argue that their paper, again, shows the need for adaptive uncertainty resolution to best utilise their modelling notation. 

\medskip\noindent
\textbf{Planning-based approaches:}
Various planing-based approaches for tackling the decision-making process under uncertainties have also been proposed in the literature~\cite{SBS:SoSYM22,Agha:IROS14,paucar2018re,martins2019alphapomdp,QosMOS,EsfahaniKM11}. Partially Observable Markov Decision Processes (POMDP) have been extensively used: (i) in the robotic domain to reason for imperfect robot actions and environment observations~\cite{Lauri:Robotics23}, (ii) in partially-observable domains for online planning in the belief space of long-endurance missions~\cite{Agha:IROS14}, (iii) to deal with partial satisficements in environment changes~\cite{paucar2018re}, (iv) the human-robot uncertainty interaction problem~\cite{martins2019alphapomdp} or (v) on the decision-making process for SASs while offering awareness of non-functional requirements' priorities at runtime through a vector-valued reward function~\cite{SBS:SoSYM22}. Despite the inherent ability of POMDP to robustify systems in the presence of uncertainty, POMDP planning is computationally intractable in the worst case. Moreover, our approach goes one step further through the separation of concern between the adaptive behavior and resolving (epistemic) uncertainty in the \UMC layer. 
Esfahani et al.~\cite{EsfahaniKM11} propose POISED, in which a ``possibility'' distribution is used for tackling the complexity of automatically making adaptation decisions under internal uncertainty. \ACRONYM goes one step further by encountering trade-offs, particularly when the probability of satisfying the system's objectives falls under a certain threshold, even without prior knowledge of the probability distribution of the monitored phenomena required by POISED.

\section{Running Example}\label{sec:running-example}

\begin{figure}
    \centering
    \includegraphics[width=0.65\columnwidth]{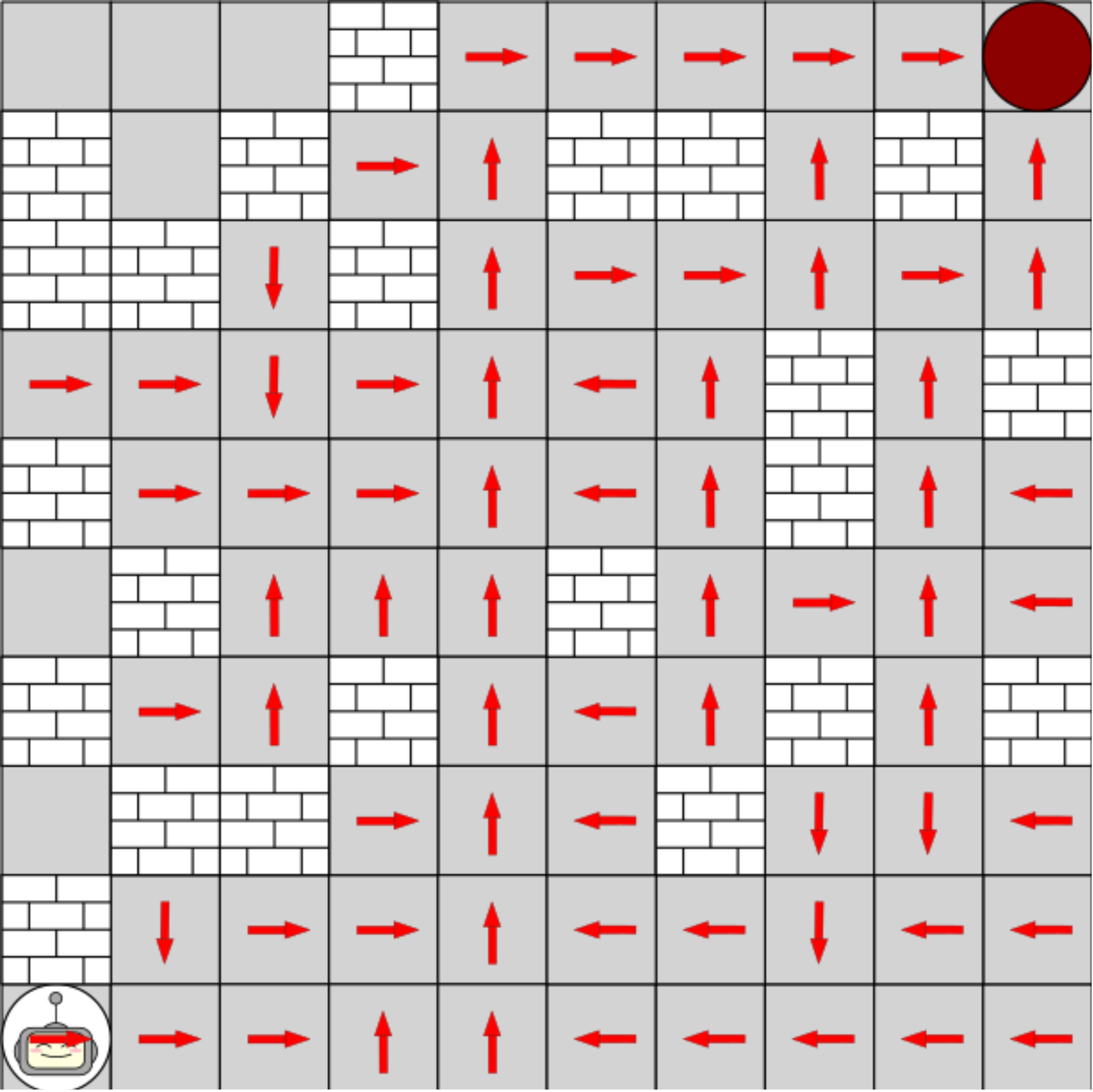}
    \vspace{-1em}
    \caption{A sample map for our running example. The robot starts in the lower left corner and its mission is to traverse to the destination in the upper right corner without crashing into static obstacles. Red arrows denote move commands.}
    \label{fig:map}
    \vspace{-2em}
\end{figure}

Throughout this paper, we use a simplified discrete mobile robot as a running example, navigating within the constraints of a known 10x10 discrete grid map. Maps of similar size have been investigated in literature~\cite{Giaquinta:NFM18}. 
The robot can move North, South, East, and West, without leaving the map. The robot's primary mission is to traverse from its initial location to a specified destination  
without crashing into static obstacles. 
Such a crash might damage the robot and is hence considered a mission failure. 
Figure~\ref{fig:map} visualises such a map, with the robot starting in the lower left corner, the destination set to the upper right corner, and walls as static obstacles. 

To facilitate the robot's movement, 
we construct a \textit{movement controller} that leverages Dijkstra's shortest path algorithm. Each move incurs a fixed cost (e.g., due to energy consumption). To discourage the robot from venturing too close to obstacles, we introduce penalties for cells near obstacles. To optimise its performance at runtime, we pre-compute moves for every conceivable position on the map, as indicated by the red arrows in Figure~\ref{fig:map}. At runtime, the movement controller selects moves based on its estimate of the robot's location 
$(\hat{x}, \hat{y})$. 

We consider that each move can have uncertain outcomes, for instance, due to wheel slip \cite{d2014hope}. However, the robot perceives all moves as successful, updating its estimate accordingly. Thus, inherent uncertainties can lead to discrepancies between the robot's estimated location and its actual location $(x, y)$ after any move.

A notable challenge arises from the controller's reliance on its estimated location when planning the next move: Eventually, this poses a risk for the robot to crash into an obstacle or not reach the specified destination.

To reduce this uncertainty, a \textit{localisation service} is available that aligns the robot's estimated location with its actual location. We assume that this service has a cost equivalent to five moves. 
The concrete problem is to determine the frequency $c$ with which the localisation service should be used concerning the estimated location 
to minimise cost while maximising the mission success rate. 
A policy for invoking the localisation service can be modelled as a function $f: (\hat{X}, \hat{Y}) \rightarrow C $ where $(\hat{X}, \hat{Y})$ represents the set of possible estimates about the robot's location and $C$ denotes a frequency for invoking the service. 

The objectives for each mission are twofold: to successfully reach the destination and to minimise cost. 
We define the following objectives accordingly: 
\begin{itemize}
    \item[(O1):] \textbf{Mission success:} The probability of reaching the destination should be maximised, and
    \item[(O2):] \textbf{Cost:} The mission cost of moving and invoking the localisation service should be minimised. 
\end{itemize}
We evaluate policies based on these objectives. Striking a balance between these objectives aims to optimise the robot's efficiency, ensuring precise navigation with resource conservation in mind.
\begin{boxA}
    We use a discrete mobile robot as a running example that navigates through a discrete map. The robot's moves suffer uncertain outcomes, and the location can only be estimated accurately by invoking a localisation service. 
    Essentially, our goal is to determine how the frequency with which the localisation service is invoked should ideally be adapted to minimise mission costs while maximizing the mission success rate.
\end{boxA}

\section{\ACRONYM}\label{sec:proposal}
In this section, we present the \ACRONYM methodology to synthesise dedicated controllers that reduce uncertainty with formal guarantees. 
First, we discuss the architecture of a system employing \ACRONYM before describing the process of \ACRONYM, including assumptions on the underlying system, synthesis of the new controller, computing formal guarantees, and finally its behaviour at runtime. 
We illustrate these points in the running example of a mobile robot introduced in the previous section. Afterwards, we elaborate on a tool that instantiates the \ACRONYM methodology to automatically generate a \UMC and synthesise policies which the stakeholders can choose from. 

\subsection{Architecture}

To separate the concerns within an adaptive or autonomous system, we propose to emphasise uncertainty reduction by establishing it as a dedicated controller, bringing it on par with controllers that are concerned with adapting the managed system's behaviour. 

In \ACRONYM, the existing notion of a \textit{change management layer} (cf.~\cite{Kramer&Magee2007}) is handled by two separate controllers, as visualised in Fig.~\ref{fig:architecture}: First, an \textit{uncertainty-aware controller} (\UUC) that employs the existing notion of adapting the managed system. This controller works distinctly with the estimates of variables, clearly incorporating uncertainty. Secondly, a novel \textit{uncertainty reduction controller} (\UMC) is solely concerned with mitigating epistemic uncertainty to aid the \UUC in its decision-making by adapting when and how the \UUC monitors the managed system and the environment.

\begin{boxA}
    With our two-layer control architecture, we propose a separation of concerns, with the \UUC reducing uncertainty in its estimates at a fixed frequency and the \UMC dedicated to adapting this frequency dynamically to optimise the system's objectives.
\end{boxA}

\noindent\textbf{Example: } In our running example, the movement controller is the \UUC, controlling the robot's movement using its estimated location and invoking the localisation service with a fixed period. 
The \UMC adapts the frequency with which the localisation service is invoked to reduce uncertainty based on the estimated location. 

\subsection{Process}

In the remainder of this section, we elaborate on the \ACRONYM methodology step-by-step from assumptions on the underlying system, to the controller synthesis, until we finally describe its runtime behaviour. 
The inputs to our automated process are:
\begin{itemize}
    \item a system model depicting the environment, UAC, managed system and uncertainty reduction services, 
    \item a range for thresholds or frequencies with which these services can be employed,
    \item a list of variables that the URC relies on for its control, and
    \item formalised system objectives;
\end{itemize}

to produce the following output: 
\begin{itemize}
    \item an extended system model with a URC, 
    \item a list of Pareto-optimal policies to trade off when to employ the uncertainty reduction services.
\end{itemize}

\subsubsection{Assumptions on the Uncertainty-Aware Controller}
First, we begin by clarifying assumptions about the \UUC and the managed system. We assume that the \UUC and the managed system are modelled as a discrete-time Markov chain (DTMC) $M$ that encodes the behaviour of the system, \emph{and} its adaptation strategies. M consists of states, a transition matrix $P$ that depicts probabilities of transitioning from one state to another and an initial state. 
Each state in $M$ is a tuple
\[
  \mathit{state}=(s,\hat{z},z),
\]
where $s$ corresponds to information fully known to the \UUC, and $\hat{z}$ is the  \UUC's estimate of the information $z$ that the controller does not know. 
Due to this uncertainty, the  \UUC makes its decisions solely based on $s$ and $\hat{z}$.
Transitions over these states, defined by a transition probability function $P$, depict the dynamic behaviour of the system. 
We further assume that the controller's estimate $\hat{z}$ can be optimised using uncertainty reduction services\footnote{Often only one such service will be available and the optimised estimate will be $z$ itself.}. The controller naively invokes the services, that is, for example with fixed frequencies, or when some variable is below a threshold. We assume that these frequencies or thresholds are encoded in a constant vector, namely $c$ such that $c$ is of dimension $n$ if $n$ uncertainty reduction services are available. 

Additionally, we assume that there are formalised objectives for the system. Typically, these refer to functional (e.g., success rate) or non-functional (e.g., costs, performance) properties. As is typical for DTMCs, a reward structure assigning rewards to selected states can be useful for modelling non-functional properties. Probabilistic model checking can be used to calculate the probability with which a property is satisfied by M or to calculate the estimated accumulated reward of M (see later steps).

\begin{boxA}
    We assume a self-adaptive system, with a controller relying on its estimate of variables suffering uncertainty to adapt the managed system. The controller naively invokes services to reduce the uncertainty in its estimate, e.g., using fixed frequencies or thresholds $c$. We assume that the system, including its controller and environment, is modelled as a DTMC. 
\end{boxA}

\noindent\textbf{Example:} For our robotic example, we construct a DTMC with PRISM~\cite{kwiatkowska2011prism} (cf. Lst.~\ref{lst:movement}). We employ PRISM's notion of modules which synchronise over labelled transitions, i.e., the transition \textit{east}, cf. ll.~\ref{line:r-east}, \ref{line:east}, \ref{line:k-east}, can be invoked only if all three modules can invoke the transition, invoking it in one single step and updating the rewards accordingly (cf. l.~\ref{line:rew-east}). In the model's first module, we depict the ground truth $z=(x, y)$ (cf. ll. \ref{line:robot-m}-\ref{line:robot-e}) resembling the robot's actual location, which changes when the robot moves.
For the second module, we employ Dijkstra's shortest path algorithm to construct the movement controller, which resembles the  \UUC, as explained in Sec.~\ref{sec:running-example}. This module (cf. ll. \ref{line:uuc-m}-\ref{line:uuc-e}) controls which move the robot should make depending on its estimated location $\hat{z} = (\hat{x}, \hat{y})$, as depicted, e.g., in l.~\ref{line:east}. 
Finally, the controller's knowledge is handled by a dedicated module that includes the estimated location, cf. ll.~\ref{line:estimate-x}-\ref{line:estimate-y}. Additionally, a $step$ counter is modelled in the Knowledge (cf. l.~\ref{line:step}), which is incremented after every move of the robot (cf. l.~\ref{line:increment-step}). A localisation service is available (cf. Sec.~\ref{sec:running-example}) that aligns the estimated location with the actual location. It is invoked if $step$ exceeds a constant limit $c$ (cf. l.~\ref{line:c}), which may have been set to, e.g., $c=2$. $step$ is reset (cf. transition $localisation$,~l.~\ref{line:update}), accordingly. In this sense, $step$ is part of the information fully known to the  \UUC, $s$. Note, that any movement or invocation of the localisation service incurs a cost (cf. ll.~\ref{line:cost-b}-\ref{line:cost-e}). We employ the objectives discussed in Sec.~\ref{sec:running-example}, accordingly. 

\begin{lstlisting}[mathescape=true, language=Prism, caption={Excerpt of PRISM code for the robot's movement.},label={lst:movement},escapechar=|]
dtmc
const int N = 9; //map size
const double p = 0.01; // probability of deviation in moves

module Robot |\label{line:robot-m}|
  x : [0..N] init 0;
  y : [0..N] init 0;
  [east] true -> |\label{line:r-east}|
    (1-3*p): (x'=min(x+1, N)) + 
    p: (y'=min(y+1, N)) + 
    p: (y'=max(y-1, 0)) + 
    p: (x'=max(x-1, 0)); 
  ...
endmodule|\label{line:robot-e}|

module Adaptation_MAPE_Controller |\label{line:uuc-m}|
  [east]  ($\hat{x}$=0) & ($\hat{y}$=0) -> true;|\label{line:east}|
  [north] ($\hat{x}$=1) & ($\hat{y}$=0) -> true;
  ...
endmodule |\label{line:uuc-e}|

const int c = 2;|\label{line:c}|
module Knowledge
  $\hat{x}$ : [0..N] init 0; //estimated position |\label{line:estimate-x}|
  $\hat{y}$ : [0..N] init 0; //estimated position|\label{line:estimate-y}|
  step : [1..10] init 1; |\label{line:step}|
  ready : Bool init true;
  [east]  ready -> ($\hat{x}$'=min($\hat{x+1}$, N)) & (ready'=false);|\label{line:k-east}|
  [north] ready -> ($\hat{y}$'=min($\hat{y+1}$, N)) & (ready'=false);
  ...
  [localisation] step>=c & !ready -> ($\hat{x}$'=x) & ($\hat{y}$'=y) & (step'=1) & (ready'=true); |\label{line:update}|
  [skip] step<c & !ready -> (step'=step+1) & (ready'=true); |\label{line:increment-step}|
endmodule

rewards "cost" |\label{line:cost-b}|
  [east] true : 1;|\label{line:rew-east}|
  [north] true : 1;
  ...
  [localisation] true : 5;
endrewards|\label{line:cost-e}|
\end{lstlisting}

\subsubsection{Synthesising the \UMC}
We propose to synthesise a dedicated controller, the \UMC, to control when the  \UUC employs its uncertainty reduction services. To this end, the \UMC adapts the vector that depicts the frequencies of updates or thresholds $c$. The \UMC performs these adaptations based on variables of the  \UUC: $s$ and $\hat{z}$. We propose that the designers select which of these variables the \UMC can use to decide how it should adapt the  \UUC\footnote{Usually, a subset of these variables will be sufficient to perform adequate adaptations.}. Note, that adaptation decisions cannot be made based on the ground truth as the ground truth is unknown to the controllers. We depict possible adaptation decisions based on the selected set of variables as a function $decision(s, \hat{z})$ for any value in the selected subset of $s$ and $\hat{z}$, accordingly, that computes a vector of dimension $n$ if $n$ uncertainty reduction services are available.
To synthesise the \UMC, we extend the DTMC such that $c$ can be set dynamically during any run, corresponding to the desired frequencies or thresholds. 
An augmented state is now a tuple 
\[
  \mathit{state_{aug}}=(s,\hat{z},z, c).
\]
where $c$ depicts the frequencies of updates or thresholds. 
We augment the transition matrix $P$ accordingly, such that the transition probability for any augmented state ($s, \hat{z}, z, *$)\footnote{With $*$ denoting a wildcard, depicting that the probability does not depend on c.} is set to the corresponding transition probability in $P$, and $c$ is set according to the \textit{decision} function, i.e., 
\[
  P'((s, \hat{z}, z, c), (s', \hat{z}', z', c')) := 
\]
\[
\begin{cases}
  P((s, \hat{z}, z), (s', \hat{z}', z')) & decision(s', \hat{z}')=c'\\
  0 & \text{otherwise}\\
\end{cases}
\] 

Therefore, any of the existing transitions, including the  \UUC's adaptations of the managed system, remain unchanged. $P'$ reflects solely the desired selection of $c$ in every transition, according to the values selected by the $decision$ function. 
Since the \textit{decision} function has not been selected yet, the concluding model $M'$ becomes a parametric DTMC (pDTMC).
Concrete values for these parameters will be determined by probabilistic model checking (see next step). One concrete instance of this function, assigning concrete values to $c$, is called a \textit{policy}.

For DTMC models in the PRISM modelling language, we provide a tool that automatically performs this augmentation to pDTMC models, using the set of variables that a decision should be based on, as well as transitions that are supposed to happen before and after the adaptations. The latter input ensures that the \UMC only adapts the \UUC when it is in a quiescent state and that the augmentation does not introduce any non-determinism. Our tool then automatically adds parameters depicting the \textit{decision} function, as well as a new module, called \textit{Uncertainty\_Reduction\_Controller} to the PRISM file making the model a pDTMC. Listing~\ref{lst:UMC} showcases such an extension: the variable $turn$ is used to depict when the  \UUC is quiescent. For practical reasons, we enforce an upper bound on $c$, which can be adapted to any value in the defined range (cf. l.~\ref{line:c_}).

\begin{boxA}
    We synthesise a \UMC that is responsible for adapting the frequencies or thresholds $c$ with which uncertainty reduction services are invoked. The adaptation decisions are modelled as parameters, resulting in a parametric DTMC. 
\end{boxA}

\noindent\textbf{Example:}
    In our example, we automatically add a module \textit{Uncertainty\_Reduction\_Controller} (cf. ll.~\ref{line:umc-m}-\ref{line:umc-e} in Lst.~\ref{lst:UMC}) to the model and move the variable $c$ to this module (cf. l.~\ref{line:c_} in Lst.~\ref{lst:UMC})\footnote{L.~\ref{line:c} in Lst.~\ref{lst:movement} is deleted, accordingly.}. Hence, the \UMC can adapt $c$. To depict quiescent states in which an adaptation is permitted, we use $turn$ (cf. l.~\ref{line:turn} in Lst.~\ref{lst:UMC}) that sequentially interrupts the system and  \UUC to allow for adaptations performed by the \UMC, i.e., when $turn=2$ (cf. l.~\ref{line:turn=2}). In our example, each simulation alternates between movements, adaptations, and finally invoking or skipping the localisation service.  We choose that potential adaptations of the \UMC depend only on the estimated location $\hat{x}$ and $\hat{y}$. Thus, the decision function depicts the \UMC's adaptations as parameters for any possible value of $\hat{x}$ and $\hat{y}$, i.e., $decision\_\hat{x}\_\hat{y}$ depicts $decision(\hat{z})$ with $\hat{z} = (\hat{x}, \hat{y})$ (cf. l~\ref{line:evolve}). We leave the parameters without concrete values since no policy has been defined yet.

\begin{lstlisting}[mathescape=true, language=Prism, caption={Excerpt of the PRISM code depicting the URC adapting $c$ based on the estimated location, automatically synthesised based on Listing~\ref{lst:movement}.\vspace{-3em}},label={lst:UMC},escapechar=|, float]
...
const int decision_0_0; //decision to invoke service at (0, 0) |\label{line:evolve}|
...
module Uncertainty_Reduction_Controller |\label{line:umc-m}|
  c : [1..10] init decision_0_0;|\label{line:c_}|
  turn : [1..3] init 1;|\label{line:turn}|
  [east] turn=1 -> (turn'=2);
  ...
  [] turn=2 & $\hat{x}$=0 & $\hat{y}$=0 -> (c'=decision_0_0) & (turn'=3);|\label{line:turn=2}|
  ...
  [localisation] turn=3 -> (turn'=1);
  [skip] turn=3 -> (turn'=1);
endmodule |\label{line:umc-e}|
\end{lstlisting}

\subsubsection{Generating Policies with Guarantees}

To find suitable policies that can satisfy the specified objectives for the pDTMC constructed in the previous step, we employ probabilistic model checking. With probabilistic model checking, we can evaluate a particular policy. If the number of possible policies is small enough, an extensive evaluation of every possible policy can be performed. 
In case the potential number of policies is very large, however, we propose to resort to a meta-heuristic approach that finds near-optimal policies. While this does not provide guarantees for optimality, invoking a probabilistic model checker such as PRISM~\cite{kwiatkowska2011prism} guarantees the near-optimal trade-offs between objectives that have been identified by the meta-heuristic search. 

\begin{boxA}
    Policies define values for the parameters and reflect the \UMC's behaviour. These policies can be found by employing probabilistic model checking of the pDTMC. If the objective space is too large, meta-heuristic search can be applied. 
\end{boxA}

\begin{figure*}
    \begin{subfigure}[t]{.45\linewidth}
        \centering
        \includegraphics[width=\linewidth]{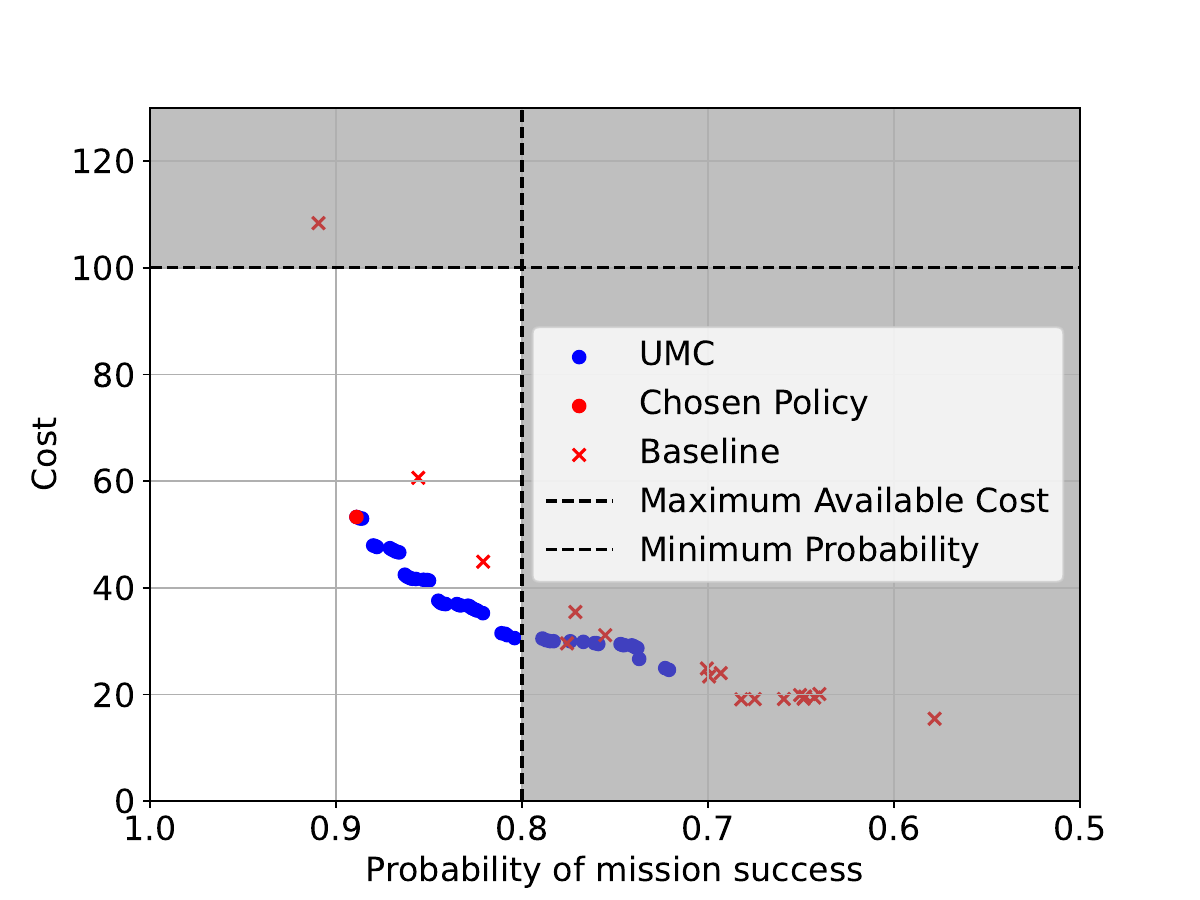}
        \caption{Pareto-front of policies generated by EvoChecker. The stakeholders determine a maximum cost of 100 and select one policy, accordingly.}
        \label{fig:pareto-front}
    \end{subfigure}
    \hspace{3em}
    \begin{subfigure}[t]{.45\linewidth}
        \centering
        \includegraphics[width=0.6\linewidth]{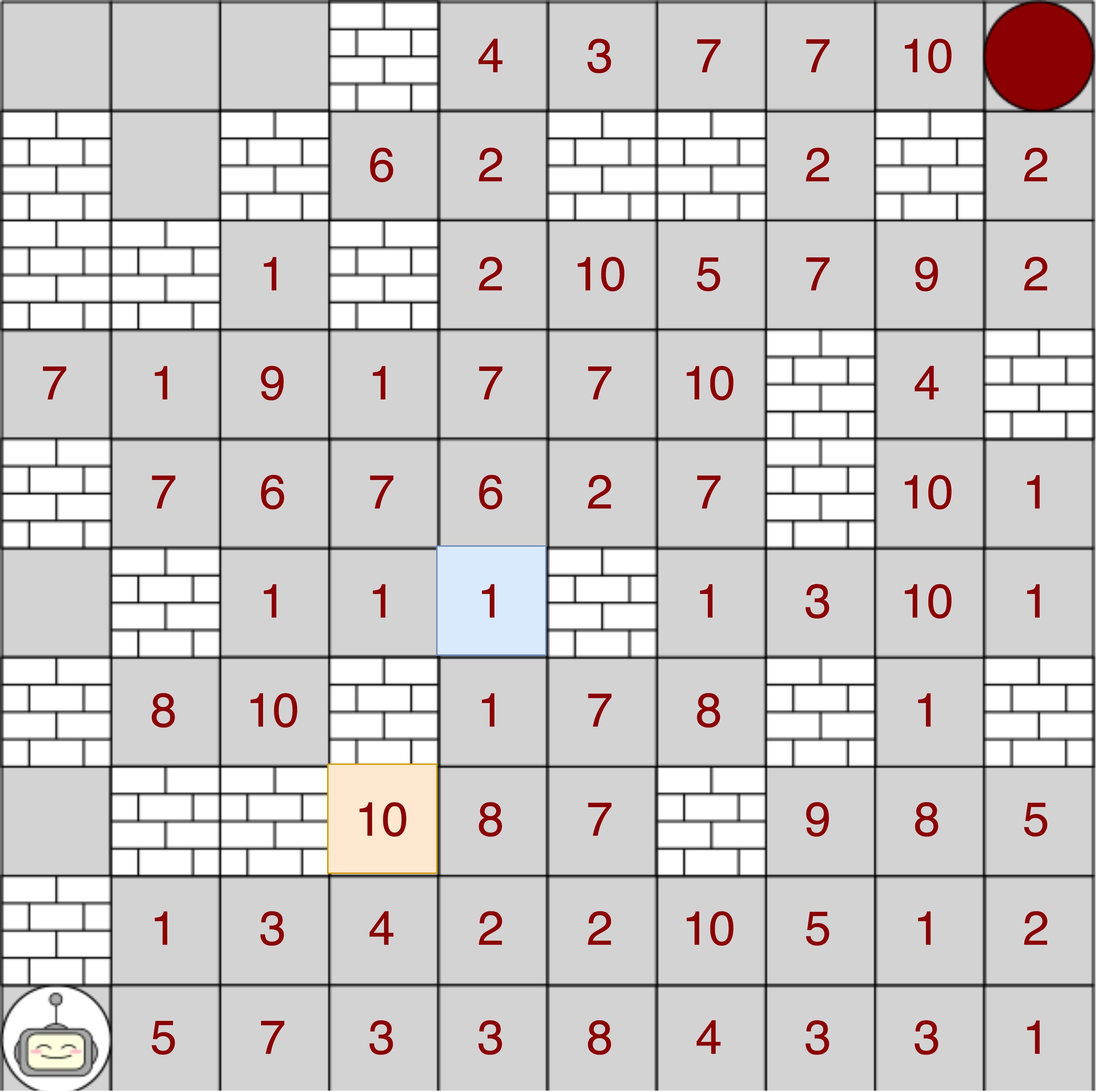}
        \caption{Extension of the map shown previously. The selected policy is visualised by numbers showcasing the intervals between localisations.}
        \label{fig:map-UMC}
        \vspace{-1em}
    \end{subfigure}
    \vspace{-1em}
    \caption{Policy selection for the robotic example.}
    \vspace{-1em}
\end{figure*}

\noindent\textbf{Example: }In our running example, each policy consists of 100 parameters, one for each possible (estimated) location on the map (100 cells in the 10x10 map). We choose to limit the maximum interval between two invocations of the localisation service to ten steps due to the amount of obstacles that are present on the maps and the resulting high likelihood of a crash. 
Since each parameter can denote any interval between 1 and 10, the number of possible policies is $10^{100}$. Due to the large search space, we resort to EvoChecker \cite{gerasimou2018synthesis}, a meta-heuristic approach to finding policies. Figure~\ref{fig:pareto-front} visualises the Pareto-front of policies that EvoChecker provides for the given map (cf. Fig.~\ref{fig:map}) and guaranteed trade-offs of the objectives. 

\subsubsection{Selecting a Policy for the \UMC}

The policies identified in the previous step form a Pareto-front concerning the specified objectives. At runtime, the stakeholders can trade off the objectives and select one of the policies of the Pareto-front accordingly. This trade-off is essential as it enables the stakeholders to prioritise the objectives, i.e., the requirements. Potential techniques to automatically select a policy, for instance with the knee method, can be used. The selected policy is deployed in the \UMC. 

\begin{boxA}
    Stakeholders select a policy from the Pareto-front by trading off objectives and prioritising requirements, accordingly.
\end{boxA}

\noindent\textbf{Example: }In our robotic example, we encounter that the robot only has limited energy available during its mission. Therefore, we only consider policies that incur a cost of less than $100$. Additionally, the stakeholders decide that the probability of a successful mission should exceed $80\%$. Fortunately, the Pareto-front contains policies satisfying these constraints. Otherwise, the stakeholders would have to compromise and prioritise one of the requirements, for example, if the map contains so many obstacles on the robot's main path to its destination that the chance of a crash is greater than usual. In our example, however, we choose the policy depicted by a red dot in Figure~\ref{fig:pareto-front} as it guarantees the highest mission success probability while staying below a cost of 100. The parameters provided by this policy are displayed in Figure~\ref{fig:map-UMC}. Every identified policy can be found in our publicly accessible online repository\footnote{\url{https://github.com/carwehlm/PARLEY/tree/main/plots/fronts}}.

\subsubsection{Enacting and Executing the Selected Policy}
At runtime, the \UMC monitors the variables chosen for its decision-making. Using the selected policy, it adapts the frequencies or thresholds $c$ in the  \UUC. 

\begin{boxA}
    The selected policy is invoked at runtime and the \UMC adapts when the uncertainty reduction services are invoked. 
\end{boxA}

\noindent\textbf{Example: }
In the robotic example, the \UMC updates the intervals between invocations of the localisation service by the  \UUC at runtime. We make the following observation: The \UMC adapts $c$ such that shorter intervals are used when the robot's planned path has an obstacle nearby, e.g., at location (4, 4) (shaded blue cell in Fig.~\ref{fig:map-UMC}). However, in locations that the robot only steps into due to deviations in its movement, such as (3, 2) (shaded orange cell in Fig.~\ref{fig:map-UMC}), the interval is set to the maximum. We assume that this is because the robot can only estimate that it is in this location when it has performed a localisation anyway. 
Thus, the \UMC helps to reduce the use of resources while maintaining a high probability of a successful mission (cf. Fig.~\ref{fig:pareto-front}).

\subsection{Automated Tool}
We instantiate the \ACRONYM methodology and provide an automated tool\footnote{\repourl} that performs the steps described in the previous subsections.
Our tool uses the following inputs: 
\begin{itemize}
    \item a PRISM file containing a DTMC depicting the managed system, environment (ground truth) and URC with its estimates,
    \item transition labels occurring \textit{before} adaptations from the \UMC,
    \item transition labels occurring \textit{after} adaptations from the \UMC, and
    \item a list of variables which are used for the \UMC's decision.
\end{itemize}
to automatically add modules to the PRISM file depicting the \UMC, as shown in Listing~\ref{lst:UMC}. Afterwards, our tool automatically calls EvoChecker and uses objectives defined in PCTL to provide a list of Pareto-optimal policies which the stakeholders can choose from. 

\section{Evaluation \label{sec:evaluation}}
	
To evaluate \ACRONYM, we formulate the following research questions that we investigate on the robotic use case outlined previously.
	
\begin{itemize}
    \item [RQ1:] \textbf{Effectiveness:} 
    How effective is \ACRONYM's adaptive uncertainty reduction in terms of achieving the system's objectives (success rate and costs) compared to a baseline that resolves uncertainty periodically?
    \item [RQ2:] \textbf{Diversity:} 
    How diverse are the policies provided by \ACRONYM to enable trading off multiple objectives compared to the baseline?
    \item [RQ3:] \textbf{Scalability:} How scalable is \ACRONYM when increasing the model size (size of the map)?

\end{itemize}

\noindent To evaluate the practicality of \ACRONYM, we formulate an additional research question:

\begin{itemize}
        \item [RQ4:] \textbf{Practicality:} 
        Does \ACRONYM work in a realistic setting of a robotic use case, and is \ACRONYM applicable to other types of systems?
\end{itemize}

\noindent \textbf{Experimental Setup for RQ1 and RQ2:} 
To evaluate the effectiveness and diversity of \ACRONYM on the robotic use case (Sec.~\ref{sec:running-example}) according to the PRISM model discussed in Sec.~\ref{sec:proposal}, we randomly generate 90 maps of size 10x10. The robot should traverse from the bottom-left to the upper-right corner of the map. Obstacles are placed randomly on the map\footnote{For each cell, we generate a random number with a standard normal distribution. If the number is outside of $[-\sigma, \sigma]$, an obstacle is placed in the cell. For every map we used, we checked that there exists a path from the robot's start to the destination.}.
Based on the 90 individual maps and their movement controllers, we synthesise 90 PRISM models, as discussed in Sec.~\ref{sec:proposal}.
In this context, uncertainty occurs due to non-deterministic outcomes of the robot's movements and it can be reduced by using a localisation service to obtain the actual location of the robot (cf. Sec.~\ref{sec:running-example}). We use the objectives defined in Sec.~\ref{sec:running-example} accordingly. 

We compare \ACRONYM to a \textit{baseline} on each of the 90 PRISM models denoting 90 \textit{scenarios} of the robotic use case. The baseline generates 10 policies that invoke the localisation service with fixed frequencies (i.e., periodically). The first policy invokes the localisation service after each movement of the robot, the second policy after every other movement, etc.\footnote{As discussed in the previous section, we set the maximum interval between two invocations of the service at 10 steps.}
The baseline policies form a Pareto-front concerning the objectives, as depicted by the red crosses in Fig.~\ref{fig:pareto-front}. 
In contrast, \ACRONYM generates policies that adapt the frequency with which the localisation service is invoked depending on the robot's estimated location. Policies found by \ACRONYM also form a Pareto-front, as depicted by the circles in Fig.~\ref{fig:pareto-front}. 

To quantify the effectiveness (RQ1), we compute the \textit{Hypervolume}~\cite{Li+2022} that measures the size of the objective space covered by the Pareto-front obtained by each approach. Hypervolume is considered a comprehensive quality indicator for the convergence, diversity, and cardinality of a solution set that is applicable if the number of objectives is rather low, which is true in our case. It can be used to compare two solution sets while a higher Hypervolume indicates a ``better'' set in terms of Pareto dominance~\cite{Li+2022}.
To quantify the diversity (RQ2), we compute the \textit{Spread}~\cite{Li+2022} of a Pareto-front obtained by each approach. Spread is a quality indicator dedicated to the diversity of solutions on a Pareto front. It is applicable to bi-objective problems and measures the distribution and uniformity of the solutions in a front. A smaller value is preferred, indicating a better distribution~\cite{Li+2022}.  
For the a-posterior analysis of the Pareto-fronts obtained by \ACRONYM and the baseline, we compute the Hypervolume and Spread concerning practically relevant solutions, which are constrained by individual requirements on the two objectives. 
Particularly, we denote with $min\_success \in \{0.6, 0.7, 0.8\}$ the minimal success rate (Objective O1 from Sec.~\ref{sec:running-example}) and with $max\_costs \in \{100, 80, 60\}$ maximal costs that are required (Objective O2 from Sec.~\ref{sec:running-example}). We consider all combinations of these two requirements, resulting in nine \textit{settings} to cover a wide range of practically relevant solutions. We depict one such setting with dotted lines in Fig.~\ref{fig:pareto-front} and only investigate policies that are within the area of acceptable policies (non-shaded area). 

To address the stochastic nature of the metaheuristic search (EvoChecker\footnote{We use EvoChecker out of the box and set a population size of 100 across 40 generations.}) in \ACRONYM, we run \ACRONYM ten times on each map. In contrast, the baseline is deterministic so that one run is sufficient. We further quantify the gain of \ACRONYM concerning Hypervolume and Spread, and test for statistical significance of the gain using Mann-Whitney U and a 95\% confidence level (p<0.05) (cf.~\cite{Arcuri+Briand2014}).

\smallskip\noindent
\textbf{Experimental Setup for RQ3:} To investigate \ACRONYM's scalability, we employ the same setup of a robotic use case described previously but scale the map size from 5 to 20 in steps of 5 to analyse how large the DTMC's state space is, how quickly PRISM can verify the objectives to evaluate a policy\footnote{ We use PRISM in version 4.7 on an M1 MacBook Pro with 16GB RAM.}, and finally how large the objective space is for \ACRONYM.

\smallskip\noindent
\textbf{Experimental Setup for RQ4:} 
We investigate RQ4 in two dimensions: We deploy \ACRONYM to \textit{ROS Gazebo}\footnote{Gazebo is a simulator for systems based on the Robotic Operating System (ROS).} to demonstrate its feasibility in a realistic setting. We further apply \ACRONYM to a self-protecting web application used in the literature~\cite{Moreno:SEAMS18} to demonstrate its applicability to a different type of system.

\subsection{RQ1: Effectiveness}\label{sec:eval:RQ1}
\begin{figure*}[bth!]
    \centering
    \vspace{-1em}
    \begin{subfigure}[t]{.5\linewidth}
      \includegraphics[width=\linewidth]{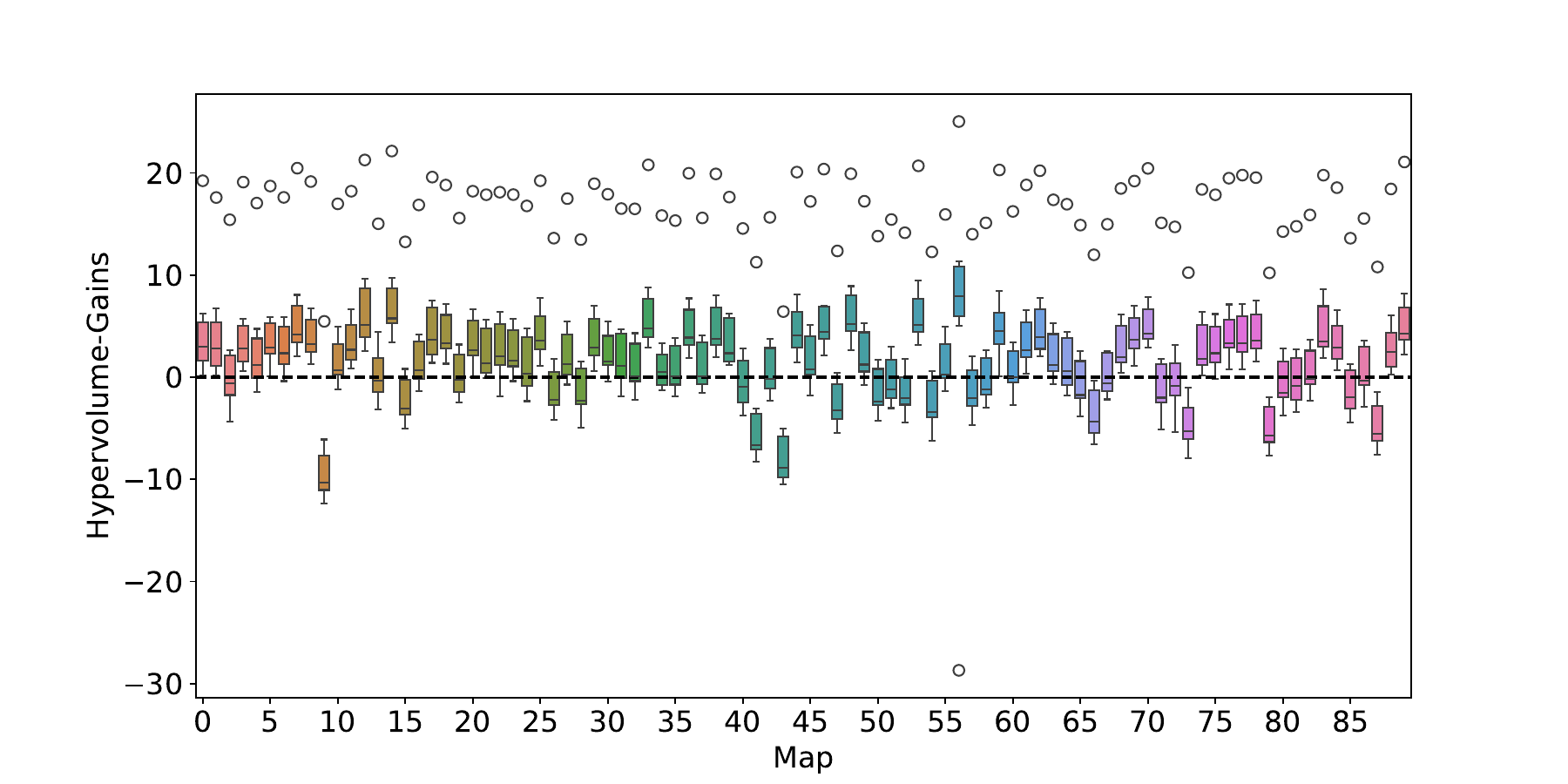}
        \caption{Gain in Hypervolume of \ACRONYM, higher is better.}
        \label{fig:box_plot-hypervolume}
    \end{subfigure}
    %\hspace{1em}
    ~
    \begin{subfigure}[t]{.5\linewidth}
        \includegraphics[width=\linewidth]{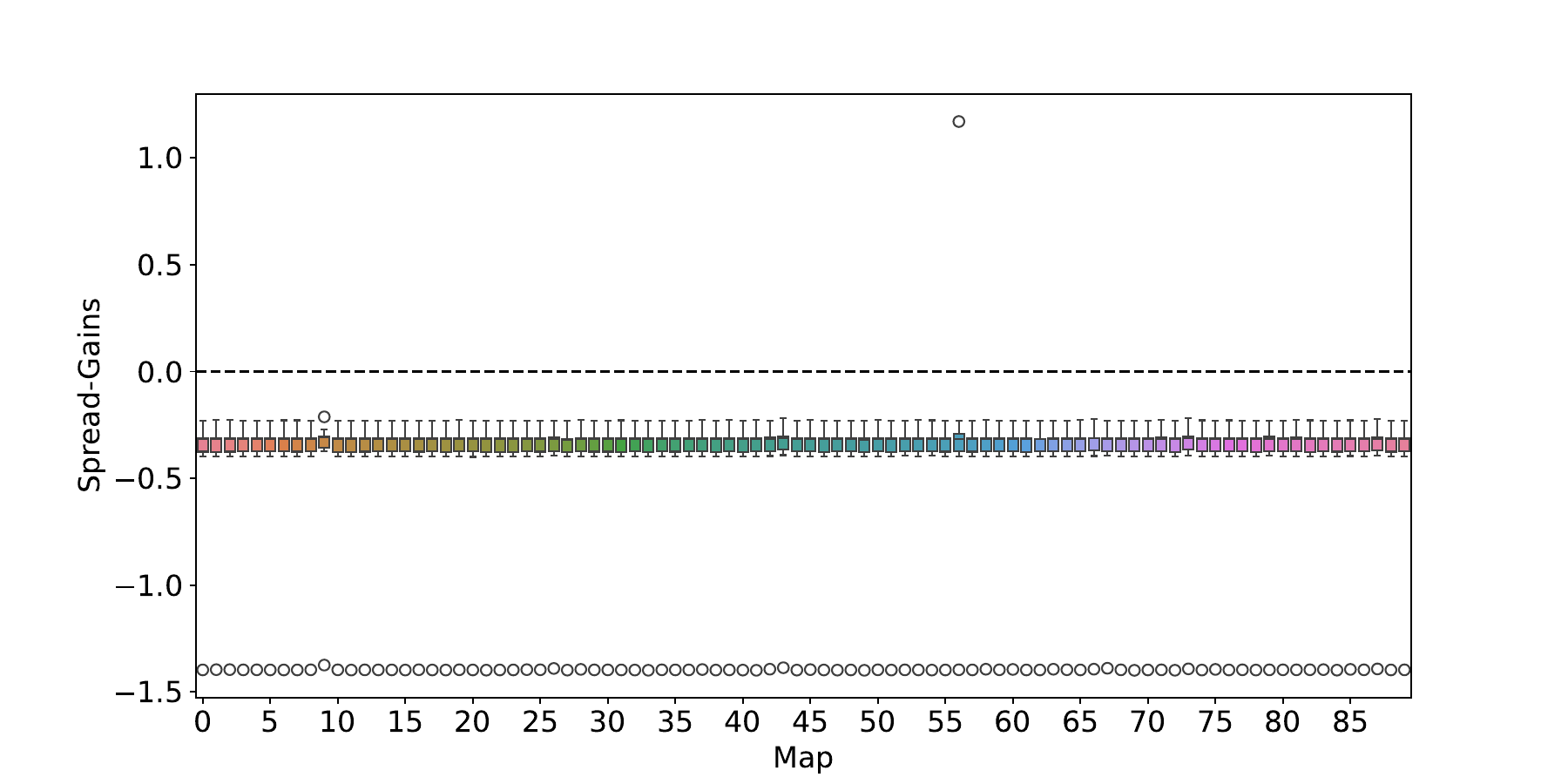}
        \caption{Gain in Spread of \ACRONYM, lower is better.}
        \label{fig:box_plot-spread}
    \end{subfigure}
    \vspace{-1em}
    \caption{Gain in (a) Hypervolume and (b) Spread of \ACRONYM compared to baseline across different maps in requirement setting $min\_success \geq 70\% $ and $ max\_cost \leq 80$.}
    \vspace{-1em}
\end{figure*}

For the first research question, we investigate if the policies provided by \ACRONYM are better, that is, closer to the optimum (minimal cost and maximum probability of a mission success), than those provided by the baseline. Table~\ref{tab:Combined} reports the results in terms of Hypervolume gains achieved by \ACRONYM over the baseline in the upper entries of each row. 
For instance, for a minimal success rate of 70\% and maximal cost of 80, \ACRONYM significantly outperforms the baseline on 48 out of 90 maps, it is significantly outperformed by the baseline in 11 maps, and the results for the remaining 31 maps are similar (i.e., no statistically significant difference). Most importantly, we can observe that \ACRONYM improves its performance over the baseline when we strengthen the requirements, that is, we increase the required minimal success rate ($min\_success$) and reduce the maximal cost ($max\_cost$). 
For strong requirements of $min\_success \geq 80\%$ and $max\_cost \leq 60$, \ACRONYM significantly outperforms the baseline in 60 out of 90 maps. In contrast, for weak requirements of $min\_success \geq 60\%$ and $max\_cost \leq 60$, the baseline performs better than \ACRONYM on more maps. This shows that for cases with weak requirements, a naive approach such as the baseline can be used, while cases with stronger requirements demand a smart approach such as \ACRONYM.
Thus, on average across all nine requirement settings (each with 90 maps), \ACRONYM provides statistically significant improvements in 405 out of 810 cases. Figure~\ref{fig:box_plot-hypervolume} visualises the results for each map for one setting ($min\_success \geq 70\% $ and $ max\_cost \leq 80$)\footnote{We provide plots for other settings in our replication package~\url{https://github.com/carwehlm/PARLEY/tree/main/plots/box-plots}
.}.

\begin{table}[tb]
    \centering
    \caption{Results for gains in Hypervolume (HV) and Spread (SP) of \ACRONYM compared to baseline in all settings, denoted by three values: the number of maps in which \ACRONYM is significantly better / significantly worse / insignificant compared to the baseline.}
    \label{tab:Combined}
    \vspace{-1em}
    \begin{tabular}{c c c c c}
        \toprule
        \multirow{2}{*}{\textbf{min\_success}} & \multirow{2}{*}{\textbf{}} & \multicolumn{3}{c}{\textbf{max\_cost}} \\
        \cmidrule(lr){3-5}
         & & \textbf{60} & \textbf{80} & \textbf{100} \\
        \midrule
        \multirow{2}{*}[1.5ex]{\textbf{60\%}} & \makecell[l]{HV\\SP} & \makecell[l]{29/34/27\\90/00/00} & \makecell[l]{36/27/27\\89/01/00} & \makecell[l]{37/22/31\\89/01/00}\\
        \midrule
        \multirow{2}{*}[1.5ex]{\textbf{70\%}} & \makecell[l]{HV\\SP} & \makecell[l]{46/11/33\\90/00/00} & \makecell[l]{48/11/31\\90/00/00} & \makecell[l]{45/11/34\\90/00/00}\\
        \midrule
        \multirow{2}{*}[1.5ex]{\textbf{80\%}} & \makecell[l]{HV\\SP} & \makecell[l]{60/12/18\\86/00/04} & \makecell[l]{54/12/24\\86/02/02} & \makecell[l]{50/11/29\\86/02/02}\\
        \bottomrule
    \end{tabular}
    \vspace{-1em}
\end{table}

\begin{boxA}
    We conclude that \ACRONYM significantly helps to achieve the system's objectives by determining when to reduce uncertainty in $50\%$ (405/810) of the cases compared to the baseline. Most importantly, it showed increased performance improvements over the baseline when strengthening the requirements. In such settings, \ACRONYM would be preferred over the baseline.
\end{boxA}

\subsection{RQ2: Diversity}
For the second research question, we investigate if the policies provided by \ACRONYM provide more diverse trade-offs than those provided by the baseline. If so, more different trade-offs along the Pareto front rather than similar trade-offs are available and can be selected by stakeholders for decision-making, which results in more diverse options for self-adaptation.
To this end, we compute the Spread (a lower value indicates a higher diversity). Table~\ref{tab:Combined} reports these results in the lower entries of each row for the nine requirement settings. 
For instance, for a minimal success rate of 70\% and maximal cost of 80, \ACRONYM significantly outperforms the baseline on all 90 maps. For all settings, \ACRONYM achieves significantly better results for at least 86 out of 90 maps, and
on average, across all nine settings, \ACRONYM provides statistically significant improvements in 796 out of 810 cases. Figure~\ref{fig:box_plot-spread} additionally visualises the gain in Spread (lower is better) achieved by \ACRONYM over the baseline for one setting  ($min\_success \geq 70\% $ and $ max\_cost \leq 80$).

\begin{boxA}
    We conclude that the policies provided by \ACRONYM are significantly more diverse than the baseline policies in $98.3\%$ (796/810) of the cases, which offers a more diverse set of trade-offs, from which stakeholders can select for self-adaptation.
\end{boxA}

\subsection{RQ3: Scalability}
In the third research question, we are concerned with the scalability of our approach. Multiple facets need to be considered here: 
\begin{itemize}
    \item \textbf{State space:} The state space of the pDTMC determines how long the evaluation of a policy with a model checker such as PRISM takes, 
    \item \textbf{Search space:} The domains of variables that are available to make a decision, as well as all possible decisions, determine the search space of possible solutions. 
\end{itemize} 
\noindent We investigate different map sizes of the running example. 
Tab.~\ref{tab:scalability} shows the results. 
\begin{table}[b]
    \vspace{-1em}
    \begin{center}
        \caption{Scalability of \ACRONYM. With larger maps, the pDTMC's state space (\#S)  increases as well as the time of model checking in PRISM for the two properties O1 and O2, and the search space of possible policies.}
        \label{tab:scalability}
        \vspace{-1em}
        \begin{tabular}{c c c c}
            \toprule
            \textbf{Map size} & \textbf{\#S} & \textbf{PRISM (s)} & \textbf{Search space}\\
            \midrule
            5x5   & 2,413 & 0.01, 0.032 & $5^{25} \approx 10^{17}$\\
            10x10 & 2,977 & 0.092, 0.067 & $10^{100}$\\
            15x15 & 6,593 & 0.313, 0.297 & $15^{225} \approx 10^{264} $\\
            20x20 & 11,276 & 0.864, 0.897 & $20^{400} \approx 10^{520}$\\
            \bottomrule
            \end{tabular}
    \end{center}
\end{table}
We can see an exponential growth in the pDTMC state space, the time to perform model checking grows accordingly. In combination with the exponential growth of the search space for possible solutions, we conclude that our approach might not scale well to larger problems.  
For a larger map (20x20) and 50 generations with a population size of 50, the model checker would be invoked 2,500 times per objective (property), resulting in a total verification time of $\approx 83$ minutes (2,500 $\times$ 1s $\times$ 2 properties). In contrast, model checking every policy with PRISM would take $\approx 10^{518}$ minutes ($20^{400}$ $\times$ 1s $\times$ 2 properties). Thus, \ACRONYM improves the scalability, which, however, can still be an issue for applying \ACRONYM at runtime in highly dynamic systems.

\begin{boxA}
    Scalability remains an issue for \ACRONYM but might be tackled by more abstract models, more computational resources, or a reduced search space for policies.
\end{boxA}

\subsection{RQ 4: Practicality}
To investigate the final research question, we first apply \ACRONYM in a realistic robotic setting. Afterwards, we investigate how \ACRONYM can be applied to the self-protecting web application.

\smallskip\noindent
\textbf{Continuous Robot:} 
Expanding from the problem investigated so far we consider a more realistic setup. Specifically, a Turtlebot3 Waffle\footnote{\url{https://www.turtlebot.com/turtlebot3/}} must travel to a destination while not crashing into obstacles or leaving the area. The system now contains a continuous robot concerning its position. The robot now also has a heading angle, $\theta$, discretised into four directions: North, East, South, and~West. 

For modelling purposes, the environment is discretised similarly to before. The environment is $21m x 21m$ discretised into $3m x 3m$ cells, resulting in a $7 x 7$ environment. 
The heading angle also suffers uncertainty, hence the controller works with an estimate that can become inaccurate over time. 
Therefore, the controller's estimates are as follows:
\begin{equation*}
      \hat{z}=(\hat{x}, \hat{y}, \hat{\theta})
\end{equation*}

\begin{figure*}
    \centering
        \begin{subfigure}{0.45\linewidth}
        \includegraphics[width=\linewidth]{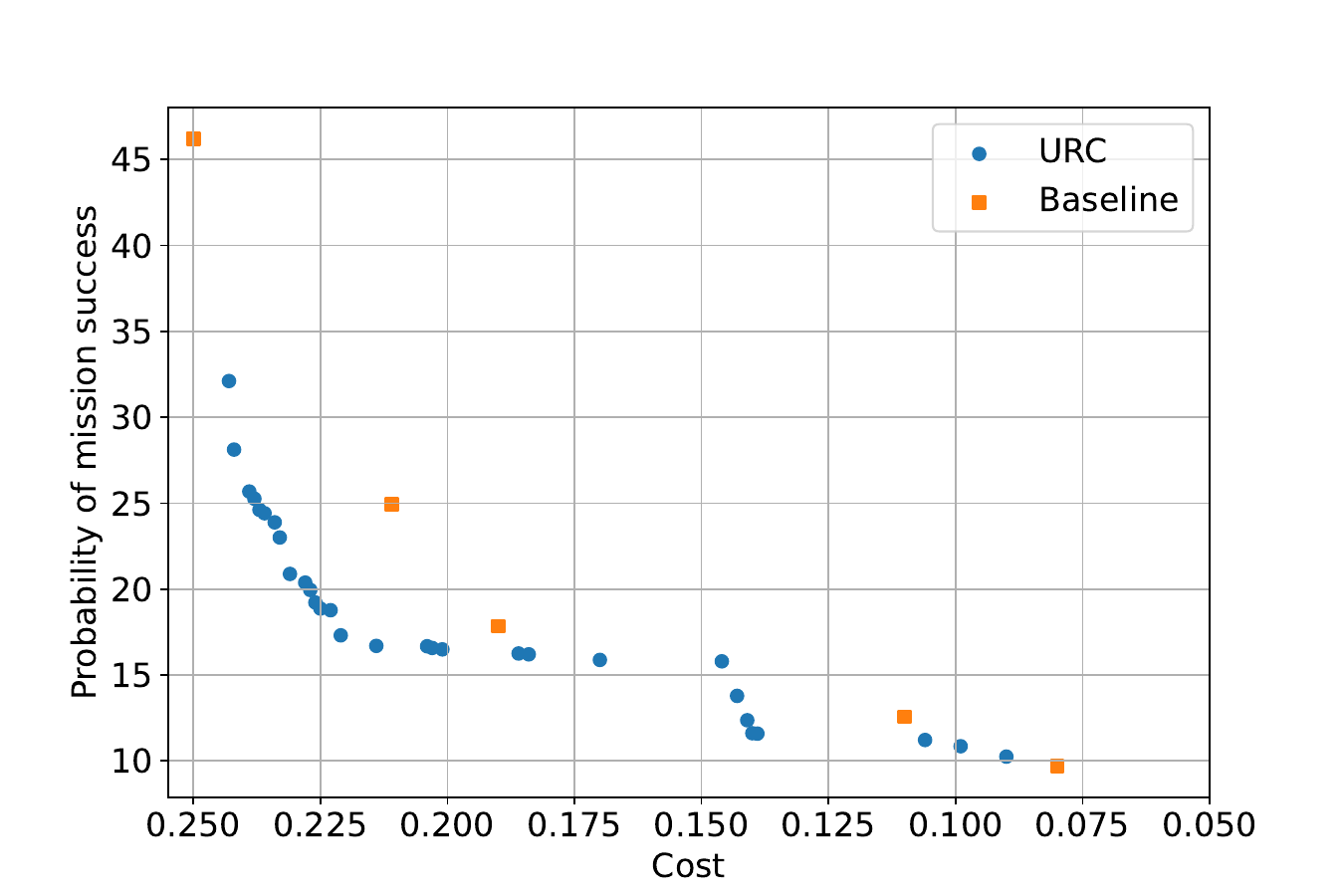}
        \caption{Continuous Robot.}
        \label{fig:gazebo}
    \end{subfigure}
    \hspace{3em}
    \begin{subfigure}{0.45\linewidth}
        \includegraphics[width=\linewidth]{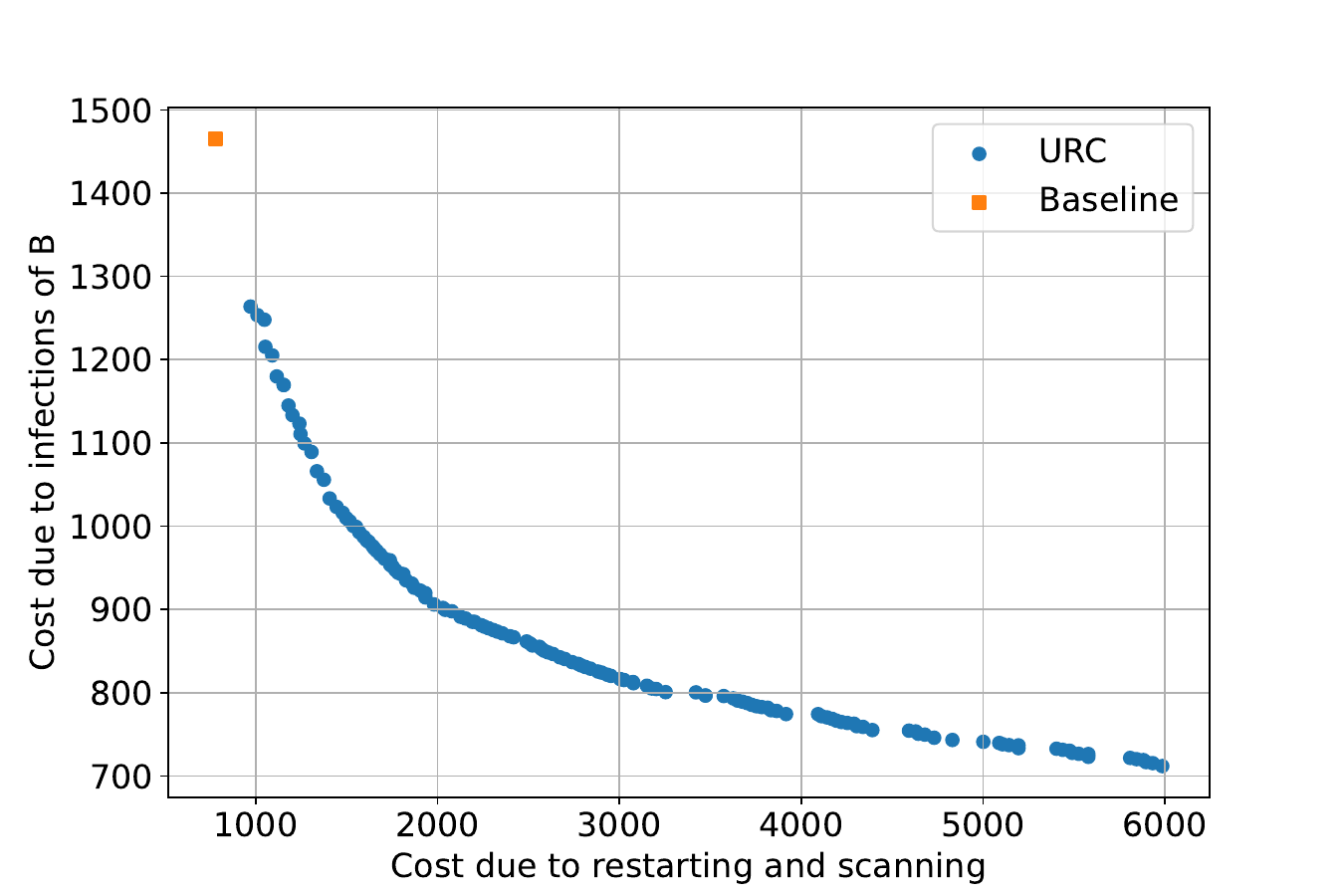}
        \caption{Web Application.}
        \label{fig:CMU}
    \end{subfigure}
    \caption{Pareto-fronts for different applications.}
\end{figure*}

The movement controller determines the shortest path as described earlier and executes the series of commands based on its estimates. However, the left wheel is slightly faulty, operating at 99\% power. This means the robot will drift over time. For this problem, the robot has no sensors but can use a localisation service similar to the running example. The system has the same objectives as the running example. We set the maximum frequency between two localisations to five given the robot's high drift and apply \ACRONYM to generate a Pareto-front of policies, visualised in Fig.~\ref{fig:gazebo}. \ACRONYM provides 30 Pareto-optimal policies compared to just five policies found in the baseline. 

We conducted simulations in Gazebo using ROS packages. To acquire the probability transitions the robot was first initialised in a cell, and chose random actions (North, East, South, West) until the robot crashed or left the environment. These traces were recorded to attain the state transitions and the corresponding probabilities. The code for running the experiments can be found online, along with a video capturing the simulation\footnote{\url{https://github.com/carwehlm/PARLEY/blob/main/turtlebot.mp4}}.

\smallskip\noindent
\textbf{Web Application:}
Moreno et al.~\cite{Moreno:SEAMS18} presented a web application containing a web server (A) and a database (B). They motivate that an attacker might infiltrate A and that the system uses an intrusion detection system (IDS) to receive alerts about such cyber-attacks. 
If the attacker is successful, A is compromised, and the attacker might go on to infiltrate B as well. In case B, too, becomes compromised, a high cost is assumed due to the sensitivity of the data in B. A controller has the options to restore A to a non-compromised version (\textit{restoreA}), restore both servers to non-compromised versions (\textit{restoreAB}), not intervene at all (\textit{NOP}), or to reduce the uncertainty in the IDS alert and determine accurately if A is compromised \textit{scanA}. Note, that while all actions except \textit{NOP} reduce uncertainty in some way, we only consider \textit{scanA} as an uncertainty reduction service as the other actions have additional implications on the system other than to solely reduce uncertainty, i.e., restoring a server. 

We modelled such a system with an extension to include dynamic costs of restoring either server A or both servers to depict that there might be times (e.g. during the night) when the server's downtime is not as expensive as at other times. Additionally, we model variability in the IDS to depict that the quality of its predictions may be dynamic, i.e., when the IDS's rule set has not been updated recently, its quality deteriorates. 

Similar to the original model, we use the IDS's prediction of an attack to determine the controller's action. However, we employ \ACRONYM to decide \textbf{when} \textit{scanA} should be invoked to reduce uncertainty, depending on the \textbf{cost of restoring the servers} and the \textbf{IDS's confidence}. In contrast to previous examples in this paper, we now determine thresholds instead of frequencies at which the uncertainty reduction service is invoked. Fig.~\ref{fig:CMU} displays the Pareto-front of policies generated by \ACRONYM and the baseline solution, which does not consider the dynamic cost or the IDS's confidence w.r.t. the cost of an infected database (x-axis) and cost of controlled actions (y-axis). \ACRONYM provides 139 Pareto-optimal policies compared to just a single one provided by the baseline. 

\smallskip\noindent
\textbf{Summary:}
Table~\ref{tab:applications} displays the main characteristics of all systems we investigated. While the robots are cyber-physical systems (CPS), the web application provides initial evidence that \ACRONYM is further applicable to another application domain. 
Thereby, with the robots we considered systems with larger search spaces (\# parameters), and with the web application we considered a system with a larger state space (\# states and \# transitions), which shows the practicality of \ACRONYM in different settings.

\begin{table}[tb]
    \begin{center}
        \caption{Key characteristics of the systems and pDTMC models used for the evaluation.}
        \label{tab:applications}
        \vspace{-1em}
        \sffamily
        \begin{tabular*}{\columnwidth}{L{0.24\columnwidth} L{0.15\columnwidth} L{0.2\columnwidth} L{0.28\columnwidth}}
            \toprule
             & \textbf{Discrete Robot} & \textbf{Continuous Robot} & \textbf{Web Application}\\
            \midrule
            \textbf{Application Domain} & \multicolumn{2}{c}{mobile robot navigation} & server infrastructure management\\
            \midrule
            \textbf{System Type} & CPS & CPS & multi-server system\\
            \midrule
            \textbf{pDTMC} & & & \\
            \textbf{\# states} & 2,977 & 512 & 5,630\\
            \textbf{\# transitions} & 4,779 & 1,070 & 11,540\\
            \textbf{\# parameters} & $10^{100}$ & $10^{7*7*4}$ & 132\\
            \bottomrule
            \end{tabular*}
    \end{center}
    \vspace{-1.5em}
\end{table}

\begin{boxA}
    We have successfully applied \ACRONYM to two further systems: a realistic use case of a real continuous robot (Turtlebot3), and a web application from literature. \ACRONYM was able to find more policies than the baseline in both applications: 6 times more for the continuous robot and 139 more for the server system. This provides initial evidence about the practicality of \ACRONYM in different domains and complexity of systems.
    
\end{boxA}

\subsection{Threats to validity}
Threats to the validity of our study are as follows: 

\noindent
\textbf{Internal:} The analysed maps for RQ1 and RQ2 pose a threat to internal validity. We generated 90 different maps randomly to mitigate this threat. Additionally, we conducted ten runs of EvoChecker for each of the maps to account for its meta-heuristic nature. For RQ3, the investigated maps might not be representative of maps of the selected sizes. The same holds for RQ4, where we merely demonstrated the applicability to other applications, but did not investigate different problem instances in these domains. We refer to the extensive literature on probabilistic model checkers, e.g.~\cite{kwiatkowska2011prism, hensel2021probabilistic}, for more information on their scalability. Currently, \ACRONYM uses PRISM, which limits its scalability. Using EvoChecker~\cite{gerasimou2018synthesis}, we started addressing this concern but still limited the map size. 
Additionally, we modelled the uncertainty reduction services as minimalistic mocks that rely on the ground truth. This will, usually, not be possible in reality. 
Further, we might have made mistakes in modelling any of the investigated systems. We reviewed the models internally and make them accessible publicly online. Finally, we used EvoChecker out of the box and did not experiment with tuning its hyperparameters.

\smallskip
\noindent
\textbf{External:} We selected three different systems from two different domains (cf. Tab.~\ref{tab:applications}) to mitigate the threat of generalisability. We further plan to investigate additional systems in the future.  

\smallskip\noindent
\textbf{Conclusion:} We used the Mann-Whitney U test for significant differences between \ACRONYM and the baseline with a $95\%$ confidence level and relied on the extensive guidelines in~\cite{Arcuri+Briand2014} to conduct our experiments to mitigate this threat. The selection of the requirement settings for RQ1 and RQ2 is, however, a threat to the conclusion validity. We mitigated this threat by considering nine different settings covering all combinations of weaker and stronger requirements for minimal success rate and maximal cost. 

\balance

\section{Conclusion and Future Work \label{sec:conclusion}}

In this work, we motivated the need for a controller dedicated to reducing epistemic uncertainty. We introduced \ACRONYM, an end-to-end methodology and architecture that relies on a pDTMC system model and synthesises an uncertainty reduction controller (\UMC), clearly separating the concerns between adapting the managed system's functionality and reducing uncertainty. For models in the PRISM language, we created a tool to automate the synthesis of a \UMC.
\ACRONYM further relies on probabilistic model checking to provide formal guarantees for the synthesised controller. Given the inherent limited scalability of model checking, we use EvoChecker, a meta-heuristic approach based on PRISM that trades off multiple objectives through Pareto-optimal solutions. 

\ACRONYM was evaluated on 90 instances of the robotic example with nine different requirement settings, and compared to a baseline concerning effectiveness and diversity.
The results show that \ACRONYM significantly outperforms the baseline in 50\% of all 810 cases in terms of effectiveness while showing improved effectiveness when strengthening the requirements.
For the diversity, \ACRONYM presented significant improvements compared to the baseline in 98.3\% of all 810 cases. Thus, \ACRONYM enables a richer and more diverse set of trade-offs, from which the stakeholders can choose for self-adaptation.
Finally, we demonstrated \ACRONYM's practicality by applying it to three different systems from two domains, including a self-protecting web application from the literature.

In future work, we aim to investigate if a tuning of EvoChecker's hyperparameters can guide the search toward even better results and if different abstractions (models) improve scalability. While we demonstrated \ACRONYM's practicality by applying it to three different systems, in future work we aim to apply \ACRONYM to more systems of other domains, such as TAS~\cite{weyns2015tele}. To this end, we plan to model the controller's estimations with probabilistic distributions. Additionally, we plan to investigate how changes in the model at runtime can be handled by refining selected policies with EvoChecker, e.g. when assumptions about the environment have changed. Our future plans also include applying \ACRONYM to other sources of uncertainty, such as those involving system goals and human interactions. A special emphasis will be put on including aleatoric uncertainty in the uncertainty reduction services, i.e., being able to express imperfect sensors and services.

\bigskip\noindent
\textbf{Data availability:} We make all models, data, and code publicly available in our reproduction package located at:\\\repourl.\\

\noindent\textbf{Acknowledgements:}
This work received funding from the Assuring Autonomy International Programme. 

\newpage
\balance
\bibliographystyle{ACM-Reference-Format}
\bibliography{bibliography,multicontrollers,uncertainty}
\end{document}